\documentclass[aps,prd,twocolumn,showpacs,eqsecnum,A4,amsmath,amssymb, nofootinbib]{revtex4-1}

\usepackage{graphicx}

\newcommand{\Dif}{{\rm D}}
\newcommand{\dd}{{\rm{d}}} % diferencial tj. stojate d s argumentem
\def \H {\mathcal{H}}

\def \B {\mathcal{B}}
\def \pul {\textstyle{\frac{1}{2}}}

\def \be {\begin{equation}}
\def \ee {\end{equation}}
\def \bea { \begin{eqnarray}}
\def \eea {\end{eqnarray}}
\newcommand{\rovno}{& = &} % rovnitko se zarovnanim pro pouztiti v eqnarray

\newcommand{\boldu}{\mbox{\boldmath$u$}} % bold u
\newcommand{\bolde}{\mbox{\boldmath$e$}} % bold e
\begin{document}

\title{ Explicit black hole solutions in higher-derivative gravity }

\author{J.~Podolsk\'y, R.~\v{S}varc}
\email{podolsky@mbox.troja.mff.cuni.cz}
\email{robert.svarc@mff.cuni.cz}
\affiliation{
  Institute of Theoretical Physics, Charles University, Prague,
  Faculty of Mathematics and Physics, V~Hole\v{s}ovi\v{c}k\'ach~2, 180~00 Prague 8, Czech Republic}
\author{V.~Pravda, A.~Pravdov\' a}
\email{pravda@math.cas.cz}
\email{pravdova@math.cas.cz}
\affiliation{
  Institute of Mathematics of the Czech Academy of Sciences,
  \v Zitn\' a 25, 115 67 Prague 1, Czech Republic}

\date{\today}
%\maketitle

\begin{abstract}
\noindent We present, in an explicit form, the metric for all
spherically symmetric Schwarzschild--Bach black holes in
Einstein--Weyl theory. In addition to the black hole mass, this
complete family of spacetimes involves a  parameter that
encodes the value of the Bach tensor on the horizon. When this
additional ``non-Schwarzschild parameter'' is set to zero the
Bach tensor vanishes everywhere and the ``Schwa--Bach''
solution reduces to the standard Schwarzschild metric of
general relativity. Compared with previous studies, which were
mainly based on numerical integration of a complicated form of
field equations, the new form of the metric enables us to
easily investigate geometrical and physical properties of these
black holes, such as specific tidal effects on test particles,
caused by the presence of the Bach tensor, as well as
fundamental thermodynamical quantities.
\end{abstract}

\pacs{04.20.Jb, 04.50.--h, 04.70.Bw, 04.70.Dy, 11.25.--w}

% 04.20.Jb Exact solutions
% 04.50.-h Higher-dimensional gravity and other theories of gravity
% 04.50.Kd Modified theories of gravity
% 04.70.Bw Classical black holes
% 04.70.Dy Quantum aspects of black holes, evaporation, thermodynamics
% 04.50.Gh Higher-dimensional black holes, black strings, and related objects
% 11.25.-w Strings and branes

\maketitle

\section{Introduction}

Einstein's general relativity, formulated about a century ago
\cite{Einstein1916}, is the most successful theory of gravity.
By predicting and correctly describing new fundamental phenomena such as black holes
\cite{Schwarzschild1916}, gravitational waves, and cosmic expansion, it has become
a cornerstone of modern theoretical physics and astronomy.
Most recent\-ly, its predictions have been confirmed by the
first direct detection of gravitational waves from a merger of
two black holes at cosmological distance.

Despite such enormous successes, it has its limitations.
As a classical field theory, it does not take quantum effects
into account. In order to understand them, with an ultimate vision
to unify general relativity with quantum theory, it is necessary to go beyond
the Einstein theory. In string and other effective theories, Einstein's gravity is extended by
higher-order terms in curvature that represent quantum corrections at high energies.
In particular, in \emph{quadratic gravity} theory, the usual Einstein--Hilbert action is generalized to include
the square of the Ricci scalar $R$ and a contraction of the Weyl tensor
$C_{abcd}$ \cite{Weyl1919,Bach1921}. In absence of matter, such an action reads
\be
S = \int \dd^4 x\, \sqrt{-g}\, \Big(
\gamma \, R  + \beta R^2 - \alpha\, C_{abcd}\, C^{abcd} \Big),
\label{actionQG}
\ee
where ${\gamma=1/G}$ ($G$ is the
Newtonian constant) and $\alpha,\beta$ are additional
parameters. The \emph{Einstein--Weyl} theory is obtained
by setting ${\beta=0}$. In this case, the field
equations  are
${\gamma \,(R_{ab} - {\pul} R\, g_{ab}) = 4 \alpha\,
B_{ab}}$, where $B_{ab}$ is the~\emph{Bach~tensor}
\be
B_{ab} \equiv \big( \nabla^c \nabla^d + {\pul} R^{cd} \big) C_{acbd}
\,, \label{defBach}
\ee
which is traceless, symmetric and conserved
(${ g^{ab}B_{ab}=0}$, ${B_{ab}=B_{ba}}$, ${{B_{ab}}^{;b}=0}$).
Taking the trace of the field equations we obtain ${ R
= 0 }$, so that they reduce to
\be
 R_{ab}=4k\, B_{ab}\,,
 \label{fieldeqsEWmod}
\ee
where ${ k \equiv \alpha\, G}$. For ${k=0}$, vacuum Einstein's equations
are immediately recovered.
Interestingly, in the case of \emph{general quadratic gravity}
(${\beta \ne0}$), it can be observed from the corresponding
field equations (see e.g. Eq.~(2)
in~\cite{PravdaPravdovaPodolskySvarc:2017}) that all solutions
to \eqref{fieldeqsEWmod} are also solutions of
\eqref{actionQG} since the trace of \eqref{fieldeqsEWmod}
implies ${R=0}$.

The field equations  \eqref{fieldeqsEWmod} form a highly
complicated system of fourth-order non-linear PDEs.
Only few non-trivial exact solutions are known.
Surprisingly, a static spherically symmetric
\emph{non-Schwarzschild black hole} has been recently identified and
discussed in \cite{LuPerkinsPopeStelle:2015}. Its metric
functions in standard coordinates are determined by
involved system of ODEs which was analyzed, e.g., in
\cite{LuPerkinsPopeStelle:2015b,HoldomRen:2017, KokkotasKonoplyaZhidenko:2017}, mainly by numerical
approaches.

In our contribution, we present an \emph{exact solution} for
such black holes in the form of \emph{explicit infinite
series}. Instead of using usual
coordinates, we express the metric in a more convenient form
conformal to type~D direct-product Kundt
geometries~\cite{PravdaPravdovaPodolskySvarc:2017}.
Higher-order corrections to the Einstein theory are
represented here by the conformally well-behaved Bach tensor. This leads
to a remarkable simplification, providing us with two compact
field equations whose solutions can be found in terms of
power series to \emph{any} order around \emph{any value} of a
radial coordinate. In addition to
mass, these black holes contain a further parameter
determining the components of the Bach tensor. By setting this additional
parameter to zero, the Schwarzschild metric
is recovered. These solutions in higher-derivative gravity can thus be called
\emph{Schwarzschild--Bach} (or Schwa--Bach) \emph{black holes}.

\section{New convenient form of \\ a black hole metric}
\label{BH metric}

For static spherically symmetric black holes, the metric
\begin{equation}
\dd s^2 = -h(\bar r)\,\dd t^2+\frac{\dd \bar r^2}{f(\bar r)}+\bar r^2(\dd \theta^2+\sin^2\theta\,\dd \phi^2) \,,
\label{Einstein-WeylBH}
\end{equation}
is commonly employed. The \emph{Schwarzschild} solution \cite{Schwarzschild1916}
is given by ${f=h=1-2m/\bar{r}}$. The metric
\eqref{Einstein-WeylBH} was also used in
\cite{LuPerkinsPopeStelle:2015} to investigate black holes in
quadratic gravity. It was demonstrated that such a class contains further
\emph{non-Schwarzschild} black hole for which ${f\not=h}$.

However, in this paper we are going to use an
\emph{alternative metric form}, namely
\be
\dd s^2 = \Omega^2(r)\big[
\,\dd \theta^2+\sin^2\theta\,\dd \phi^2 -2\,\dd u\,\dd r+{\cal H}(r)\,\dd u^2 \,\big].
\label{BHmetric}
\ee
This is related to the metric (\ref{Einstein-WeylBH}) via the transformation
\begin{equation}
\bar{r} = \Omega(r)\,, \qquad
  t = u - {\textstyle\int}\, {\H(r)}^{-1}\dd r \,,
  \label{to static}
\end{equation}
and the new metric functions $\Omega$, $\H$ are related to
 $f$, $h$ as
\be
  h({\bar r}) = -\Omega^2\, \H\,,\qquad
  f({\bar r}) = -(\Omega'/\Omega)^2\, \H \,,
  \label{rcehf}
\ee where $\Omega'$ denotes the derivative of $\Omega$ with
respect to $r$.

The \emph{Killing horizon} associated with ${\partial_t=\partial_u}$ is
located at $r_h$ such that
\be
\H \big|_{r=r_h}=0\,,
  \label{horizon}
\ee
and, due to \eqref{rcehf}, also ${h({\bar r_h})=0=f({\bar r_h})}$.
This is unchanged under the \emph{time-scaling freedom} ${t\to \sigma^{-1}\, t}$ implying ${h\to \sigma^2\, h}$,
which can be used, e.g., to set ${h = 1}$ at spatial infinity for asymptotically flat solutions.

The metric \eqref{BHmetric}, writtten as ${\,\dd s^2 =
\Omega^2(r)\,\dd s^2_{\hbox{\tiny Kundt}}\,}$,
is \emph{conformal to}  ${\dd s^2_{\hbox{\tiny Kundt}}}$
which belongs to the class of \emph{Kundt geometries}
\cite{Stephanietal:2003, GriffithsPodolsky:2009} (in fact,
to a subclass that is the direct-product of two
2-spaces~\cite[Chap.~7]{GriffithsPodolsky:2009}).

\section{The field equations}
\label{derivingFE}

The conformal approach to investigating black holes, based
on  the metric \eqref{BHmetric}, is very convenient since it
enables us to evaluate the Ricci and Bach tensors  from the
corresponding tensors of the simpler metric ${\dd
s^2_{\hbox{\tiny Kundt}}}$. In particular, the Bach tensor is
given by ${B_{ab} = \Omega^{-2}\,B_{ab}^{\hbox{\tiny Kundt}}}$.
A direct calculation yields three non-trivial components
of the field equations \eqref{fieldeqsEWmod} for the metric
functions $\Omega(r)$ and ${\cal H}(r)$.
By employing the Bianchi identities, it can be shown
\cite{PodolskySvarcPravdaPravdova:2018b}
that they reduce to two ODEs
\begin{align}
\Omega\Omega''-2{\Omega'}^2 = &\ \tfrac{1}{3}k\, \B_1 \H^{-1} \,, \label{Eq1}\\
\Omega\Omega'{\cal H}'+3\Omega'^2{\cal H}+\Omega^2
 = &\ \tfrac{1}{3}k \,\B_2  \,, \label{Eq2}
\end{align}
where \emph{2 independent components of the Bach tensor} are
\be
\B_1 \equiv \H \H''''\,, \qquad \B_2 \equiv \H'\H'''-\tfrac{1}{2}{\H''}^2
+2\,.
\label{B2}
\ee
This system is considerably simpler than
the previously used equations for the metric
\eqref{Einstein-WeylBH}, see e.g. \cite{KokkotasKonoplyaZhidenko:2017}.
Moreover, Eqs. \eqref{Eq1}, \eqref{Eq2} form an \emph{autonomous system}
(they do not explicitly depend on the variable $r$) which is essential for
finding their solution in the form \eqref{rozvoj} below.

Recall that the trace of \eqref{fieldeqsEWmod} gives  ${R=0}$,
which reads
\begin{equation}
{\cal H}\Omega''+{\cal H}'\Omega'+{\textstyle \frac{1}{6}} ({\cal H}''+2)\Omega = 0 \,.
 \label{trace}
\end{equation}
In fact, this equation is obtained by subtracting \eqref{Eq1}
multiplied by $\H'$ from the derivative of \eqref{Eq2}.

For a geometrical/physical interpretation, let us evaluate the
Bach and Weyl \emph{scalar curvature invariants}:
\begin{align}
  B_{ab}\, B^{ab} &=  \tfrac{1}{72}\,\Omega^{-8}\big[(\B_1)^2
+2(\B_1+\B_2 )^2\big] \,,\label{invB}\\
 C_{abcd}\, C^{abcd} &=  \tfrac{1}{3}\,\Omega^{-4}\,({\cal H}'' +2)^2 \,. \label{invC}
\end{align}
In fact, ${ B_{ab}=0}$  if (and only if) ${B_{ab}\, B^{ab}
=0}$. Moreover, ${ C_{abcd}\, C^{abcd}=0}$ implies ${B_{ab}
=0}$. Notice also from \eqref{B2}, \eqref{horizon} that
$\B_1$ \emph{always vanishes on the horizon}.
Based on the invariant \eqref{invB}, there are thus \emph{two
geometrically distinct classes of solutions} to (\ref{Eq1}),
(\ref{Eq2}), depending on the Bach tensor. The first
corresponds to ${B_{ab}=0}$, while the involved
second case arises when ${B_{ab}\ne0}$.

\section{Vanishing Bach tensor: uniqueness of
Schwarzschild} \label{integration:Schw}

In the case ${\B_1=0=\B_2}$, using a coordinate freedom ${r \to
\lambda\,r+\nu}$, ${u \to\lambda^{-1}u}$ of the metric
(\ref{BHmetric}), the complete solution of
Eqs.~(\ref{Eq1})--(\ref{B2}) is
\begin{equation}
\Omega(r)=-\frac{1}{r}\,,\qquad
\H(r) = -r^2-2m\, r^3   \,.
\label{Schw}
\end{equation}
This is the Schwarzschild solution, since \eqref{to static}, \eqref{rcehf} give
${r=-1/\bar{r}}$, $f{(\bar{r}) = 1-2m/\bar{r}= h(\bar{r})}$,
where ${\bar{r}>0}$ corresponds to
${r<0}$ ($r$ increases with ${\bar{r}}$).
The \emph{Schwarzschild black hole is thus the only possible
solution with vanishing Bach tensor}, in accordance with Birkhoff's theorem.

\section{Non-vanishing Bach tensor: general
Schwarzschild--Bach} \label{integration:nonSchw}

With ${\B_1, \B_2 \ne0}$, the system \eqref{Eq1}, \eqref{Eq2}
of non-linear field equations is
coupled in a non-trivial way. However, it is autonomous, so that its solutions can be found as
\emph{expansions in the powers of $r$ around any fixed value} ${r_0}$,
\be
\Omega(r)  = \Delta^n \,
\sum_{i=0}^\infty a_i \,\Delta^{i}\,,\qquad
\H(r)      = \Delta^p \,
\sum_{i=0}^\infty c_i \,\Delta^{i}\,,
\label{rozvoj}
\ee
where ${\Delta\equiv r-r_0}$. Inserting  the series
\eqref{rozvoj} with ${n,p\in \mathbb{R}}$ into Eqs. \eqref{Eq1},
\eqref{Eq2}, \eqref{trace}, it can be shown
\cite{PodolskySvarcPravdaPravdova:2018b} that the dominant
powers of $\Delta$ imply specific restrictions such that
\emph{only four classes of solutions of the form \eqref{rozvoj}
are allowed}, namely
${[n,p]=[-1,2]}$,
${[0,1]}$,
${[0,0]}$,
${[1,0]}$.
We have proved
\cite{PodolskySvarcPravdaPravdova:2018b} that the only
solution in the class  ${[-1,2]}$ is the
Schwarzchild black hole \eqref{Schw},
while the class ${[1,0]}$ is equivalent to the peculiar
${(s,t)=(2,2)}$ class of \cite{Stelle:1978,LuPerkinsPopeStelle:2015b}. The
Schwarzschild--Bach black hole is contained in the classes
${[0,1]}$ and ${[0,0]}$.

\subsection{Class ${[0,1]}$: Schwa--Bach black hole\\
expressed around the horizon~$r_h$}

In general, $r_0$ in $\Delta$ of expansions \eqref{rozvoj} can be any constant.
However, in the ${[0,1]}$ class, $r_0$ is the root of~$\H$, and thus the horizon $r_h$,
see Eq.~\eqref{horizon}. A lengthy analysis shows
\cite{PodolskySvarcPravdaPravdova:2018b} that this class of
solutions of the Einstein--Weyl/quadratic gravity
includes \emph{non-Schwarzchild}  black holes with ${B_{ab}\ne 0}$.
Their explicit form \eqref{rozvoj} is
\bea
\Omega(r) \rovno -\frac{1}{r}
-\frac{b}{r_h}\sum_{i=1}^\infty\alpha_i\Big(1-\frac{r}{r_h}\Big)^i , \label{Omega_[0,1]}\\
\mathcal{H}(r) \rovno (r-r_h)\bigg[  \frac{r^2}{r_h}
+3b\,r_h\sum_{i=1}^\infty
\gamma_i\Big(\frac{r}{r_h}-1\Big)^i\,\bigg] ,
\label{H_[0,1]}
\eea
where the initial coefficients are
\begin{equation}
\alpha_1=1 \,, \quad \gamma_1=1\,, \quad
\gamma_2 = \frac{1}{3}\Big(4-\frac{1}{2kr_h^2}+3b\Big) \,,
\label{alphasgammainitial_[0,1]}
\end{equation}
and $\alpha_l, \gamma_{l+1}$ for ${l \ge 2}$ are given by the
recurrent relations
\begin{align}
&\alpha_{l}= \, \frac{1}{l^2}\Big[\alpha_{l-1}\big(2l^2-2l+1\big)-\alpha_{l-2}(l-1)^2 \nonumber\\
&\ \qquad -3\sum_{i=1}^{l}(-1)^i\,\gamma_i\,(1+b\,\alpha_{l-i})\big(l(l-i)+\tfrac{1}{6}i(i+1)\big)\Big],
   \nonumber\\
&\gamma_{l+1}= \, \frac{(-1)^{l}}{kr_h^2\,(l+2)(l+1)l(l-1)} \label{alphasgammasgeneral_[0,1]}\\
&\ \qquad \times\,\sum_{i=0}^{l-1}
   \big(\alpha_i+\alpha_{l-i}(1+b\,\alpha_i) \big)(l-i)(l-1-3i) \,,
   \nonumber
\end{align}
(with ${\alpha_0\equiv 0}$) so that
${ \alpha_2 = 2+\frac{1}{8kr_h^2}+b}$,
${ \gamma_3 = \frac{1}{96k^2r_h^4}}$ etc.

	\begin{figure}[t!]
	\includegraphics[scale=0.4]{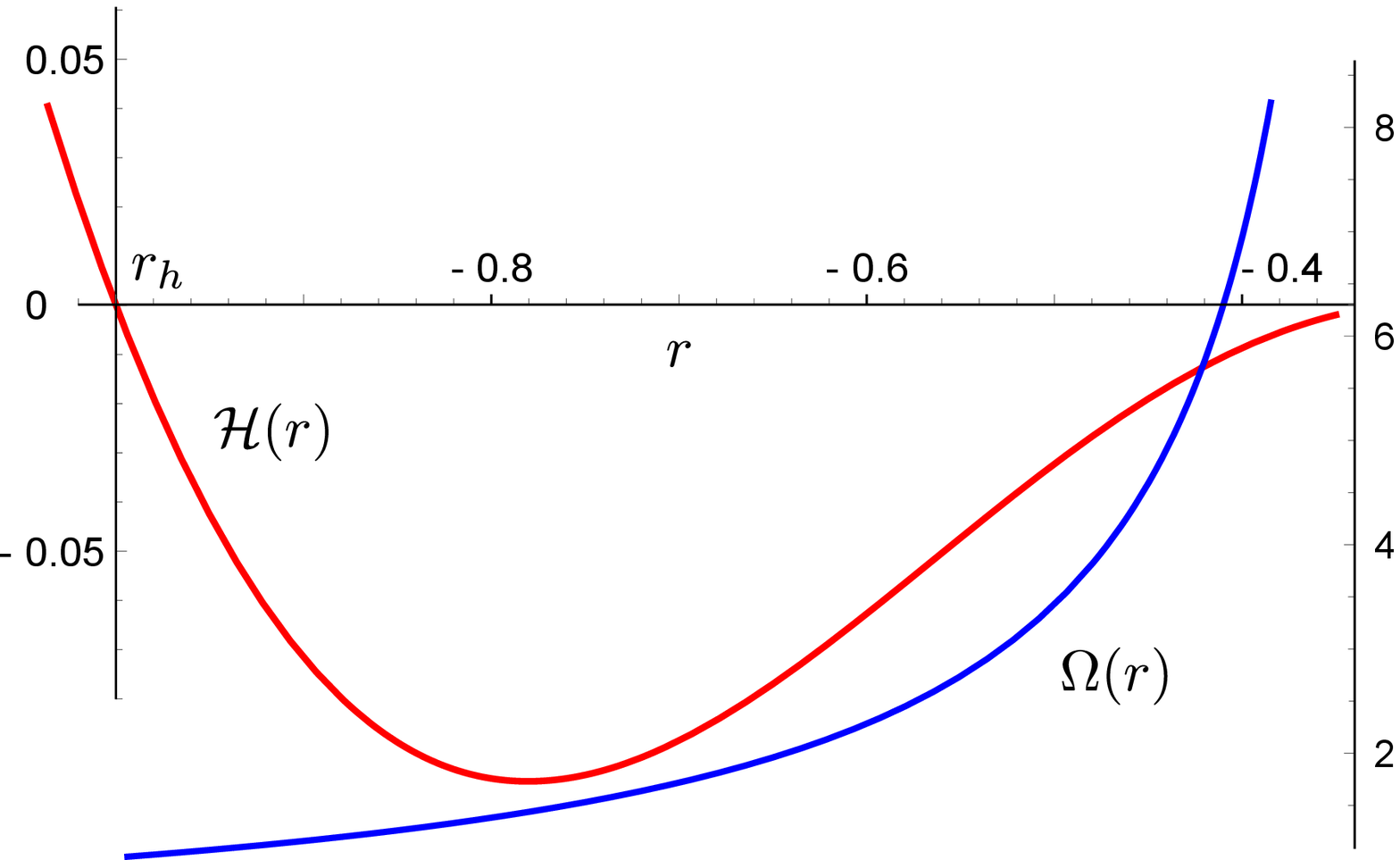}
		\caption{\label{fig:1}
            The functions  ${\cal
            H}(r)$ and $\Omega(r)$ for the Schwa--Bach black hole in the form \eqref{BHmetric}.
            The first 20 terms in \eqref{H_[0,1]} for ${\cal H}$ agree with a numerical solution with precision $10^{-4}$, and the first 40 terms in \eqref{Omega_[0,1]} for  $\Omega$ agree with precision $10^{-5}$ on ${[-1, -0.5]}$.
            The horizon is at ${r_{h}=-1}$, and  ${k=0.5}$, ${b=0.3633018769168}$, which are the same values as in~\cite{LuPerkinsPopeStelle:2015}.
            }
    \vspace{6mm}
	\includegraphics[scale=0.4]{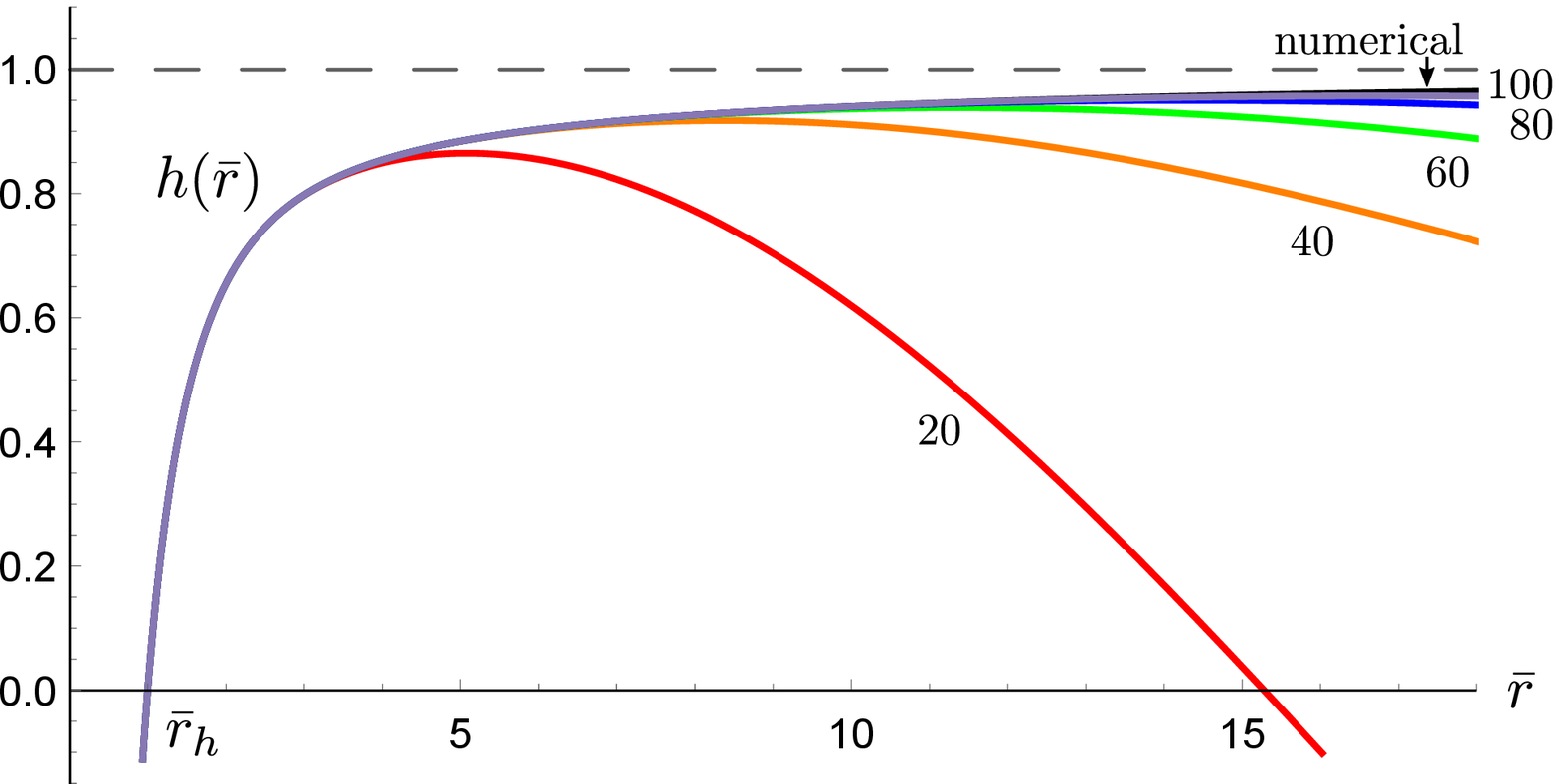}
		\caption{\label{fig:2}To demonstrate the rapid convergence in the near-horizon region, we plot the
            function $h(\bar r)$ of the metric \eqref{Einstein-WeylBH} expressed using \eqref{rcehf}.
            The first 20 (red), 40 (orange), 60 (green), 80 (blue), and 100 (violet) terms in the series
            are compared with the numerical solution of \cite{LuPerkinsPopeStelle:2015} (black). The horizon is located at ${{\bar r}_h=1}$ (that is ${r_{h}=-1}$). The scaling freedom with ${\sigma^2 \approx 2.18}$ has been used to obtain ${h \to 1}$ asymptotically. }
    %\vspace{-2mm}
 	\end{figure}

This family of spherically symmetric black holes
depends on \emph{two parameters} with a clear
interpretation:
\begin{itemize}
\item The parameter $r_h$ identifies the \emph{horizon
    position}. Clearly, ${r=r_h}$ is
     the root of $\H$ given by \eqref{H_[0,1]}.
\item The dimensionless \emph{Bach parameter} $b$
    \emph{distinguishes} the Schwarzschild solution
    (${b=0}$) from the  more general non-Schwarzschild
    (Schwa--Bach) black hole with non-zero Bach
    tensor (${b\ne0}$).
\end{itemize}
Indeed, setting ${b=0}$, the solution \eqref{Omega_[0,1]},
\eqref{H_[0,1]} reduces to
\eqref{Schw}, the Schwarzschild solution (its horizon is given by
 ${r_h=-\frac{1}{2m}}$, where $m$ is the black hole mass).
Moreover, we have chosen the new parameter $b$ to \emph{determine the value of the Bach tensor
\eqref{B2} on the horizon} $r_h$, namely
\be
\B_1(r_h) = 0\,,\qquad
\B_2(r_h) = -\frac{3}{kr_h^2}\,b \,.
\label{bonhorizon}
\ee
The invariants \eqref{invB} and \eqref{invC} are
${B_{ab}\,B^{ab}(r_h) = \frac{r_h^4}{4 k^2}\,b^2}$ and
${C_{abcd}\, C^{abcd}(r_h) = 12\,r_h^4\,(1+b)^2}$, respectively.

The behavior of the metric functions $\H$ and $\Omega$ given by \eqref{Omega_[0,1]}--\eqref{alphasgammasgeneral_[0,1]}
is shown on Fig.~\ref{fig:1} for a special value of $b$ when the Bach tensor approaches zero
for large ${\bar{r}\equiv\Omega(r)}$, see Fig.~\ref{fig:3}.
Close to the horizon, the series rapidly converge to the numerical solution of \cite{LuPerkinsPopeStelle:2015}. This can be seen in Fig.~\ref{fig:2},
where, using the parametric plot and \eqref{rcehf}, the function $h(\bar r)$ of the metric \eqref{Einstein-WeylBH} is expressed from $\Omega$ and $\H$.

\subsection{Class ${[0,0]}$: Schwa--Bach black hole\\
 expressed around any point ${r_0\ne r_h}$}

In this case, the solution to Eqs.~\eqref{Eq1}, \eqref{Eq2}
of the form \eqref{rozvoj} with ${n=0=p}$ is given by the Taylor expansions,
where ${a_0, a_1, c_0, c_1, c_2}$ are five free parameters,
\begin{align}
c_3 =&\ \frac{1}{6kc_1}\big[3a_0(a_0+a_1c_1)+9a_1^2c_0+2k(c_2^2-1)\big]\,,
\nonumber
\end{align}
\begin{align}
a_{l+1} =&\ \frac{-1}{l(l+1)\,c_0}\,\Big[\tfrac{1}{3}\,a_{l-1}\nonumber\\
&\ +\,\sum^{l+1}_{i=1}
     c_i\,a_{l+1-i}\,\big(l(l+1-i)+\tfrac{1}{6}i(i-1)\big)\Big]\,,
\nonumber\\
c_{l+3} =&\ \frac{3}{k\,(l+3)(l+2)(l+1)l}\label{[0,0]initcondc}\\
&\ \times\sum^{l}_{i=0}
a_i \,a_{l+1-i}(l+1-i)(l-3i)\,,
\nonumber
\end{align}
for any ${l\ge 1}$, see
\cite{PodolskySvarcPravdaPravdova:2018b}. This is a large class
of solutions with non-trivial Bach tensor.
To identify the Schwa--Bach black hole
\eqref{Omega_[0,1]}, \eqref{H_[0,1]}, previously expressed around the
horizon $r_h$ in the class ${[0,1]}$, we have to uniquely
determine the five free parameters by evaluating
the functions \eqref{Omega_[0,1]}, \eqref{H_[0,1]} and their
derivatives at ${r=r_0}$. Interestingly, for ${b=0}$, the coefficients $a_i$ form a geometrical series,
$r_0$ disappears, and the metric functions simplify to
the Schwarzschild solution in the form
\eqref{Schw} with ${2m=-1/r_h}$.
For ${B_{ab}=0}$, both classes ${[0,0]}$ and ${[0,1]}$
thus reduce to the Schwarzchild black hole.
Recall that the parameter $r_0$ in the class ${[0,1]}$
\emph{equals}~$r_h$, while ${r_0\ne r_h}$ can be chosen
\emph{arbitrarily} in the class ${[0,0]}$.

Since, in general, ${\B_1(r_0), \B_2(r_0)}$ are independent, the $[0,0]$ class admits one
more parameter than the Schwa-Bach black hole and thus it
is a \emph{larger family} of solutions.
Moreover, the  power  series \eqref{rozvoj} with
integer exponents transforms in some cases to  series
with \emph{non-integer} exponents in the usual
coordinate ${\bar r}$. For example, a new class $(w,t)=(4/3,0)$
in the notation of \cite{LuPerkinsPopeStelle:2015b} also
belongs to our $[0,0]$ class, see
\cite{PodolskySvarcPravdaPravdova:2018b}.

\section{Observable effects caused by the Schwa--Bach black hole}

The two independent parts $\B_1, \B_2$ of the Bach invariant
\eqref{invB} can be observed via a
\emph{specific influence on test particles}, namely
their \emph{relative motion} described by the equation of
geodesic deviation \cite{PodolskySvarc:2012}. To
obtain measurable information, we project it
onto an orthonormal frame associated with an
\emph{initially static observer} (${\dot{r}=0}$, ${\dot{\theta}=0=\dot{\phi}}$),
namely ${\bolde_{(0)}=\boldu=
\dot{u}\,\partial_u}$,
$\bolde_{(1)}=-\dot{u}\,(\partial_u+\H\,\partial_r)=
-\H\,\Omega'\,\dot{u}\,\partial_{\bar r}$,
${\bolde_{(2)}=\Omega^{-1}\partial_\theta}$,
${\bolde_{(3)}=(\Omega\sin\theta)^{-1}\partial_\phi}$. Indeed,
${\bolde_{(a)}\cdot\bolde_{(b)}=\eta_{ab}}$, and the
normalization of the observer's
velocity $\boldu$ implies ${\Omega^2\H\,\dot{u}^2=-1}$. Denoting the
\emph{relative position} of two particles as
${Z^{(a)} \equiv {e^{(a)}}_{\!\!\mu}\,Z^\mu}$, and their \emph{mutual
acceleration} as ${\ddot Z^{(a)} \equiv
{e^{(a)}}_{\!\!\mu}\,\frac{\Dif^2 Z^\mu}{\dd\, \tau^2} }$, we obtain
\begin{align}
\ddot{Z}^{(1)} = & \hspace{5.6mm} \frac{1}{6} \frac{{\cal H}''+2}{\Omega^2}\,Z^{(1)}
-\frac{k}{3}\,\frac{\B_1+\B_2}{\Omega^4}\,Z^{(1)} \,, \label{InvGeoDevBH1r0}\\
\ddot{Z}^{(i)} = & - \frac{1}{12}\frac{{\cal H}''+2}{\Omega^2}\,Z^{(i)}
-\frac{k}{6}\,\frac{\B_1}{\Omega^4}\,\,Z^{(i)} \,, \label{InvGeoDevBHir0}
\end{align}
where ${i=2,3}$.
There is the classical \emph{Newtonian tidal
deformation} caused by the \emph{Weyl curvature} proportional
to ${({\cal H}''+2)\,\Omega^{-2}}$, i.e., square root of the
invariant \eqref{invC}. The Schwa--Bach black hole
causes \emph{two additional efects} due to the \emph{Bach
tensor}. The first is observed in the \emph{transverse}
components of the acceleration \eqref{InvGeoDevBHir0} along~$\partial_\theta, \partial_\phi$,
while the second occurs in the \emph{radial}
component \eqref{InvGeoDevBH1r0} along $\partial_{\bar r}$.
Their amplitudes are given by $\B_1, \B_2$ defined in \eqref{B2}. Interestingly, \emph{on the horizon}
there is \emph{only the radial} effect caused by ${\B_2}$ since
${\B_1(r_h)=0}$, see\ \eqref{bonhorizon}. It can also be proven
\cite{PodolskySvarcPravdaPravdova:2018b} that $\B_1, \B_2$
\emph{cannot mimic} the Newtonian tidal effect, i.e., cannot
be ``incorporated'' into the first terms in
\eqref{InvGeoDevBH1r0}, \eqref{InvGeoDevBHir0}. Therefore, by
detecting free fall of a set of test particles \emph{it is
possible to distinguish} the pure Schwarzschild from the
general Schwa--Bach geometry.

	\begin{figure}[h!]
\includegraphics[scale=0.4]{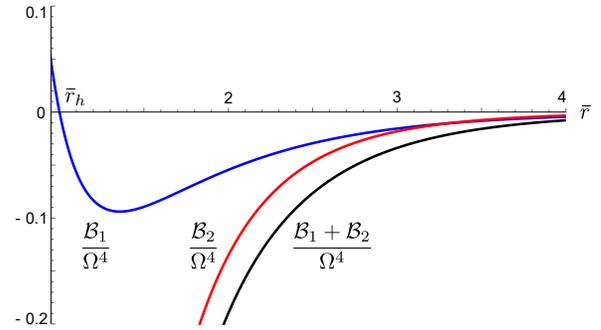}
		\caption{\label{fig:3}The Bach tensor components \eqref{B2}, entering \eqref{InvGeoDevBH1r0} and \eqref{InvGeoDevBHir0}, as functions of ${\bar r}$. For the special value of $b$ as in Fig.~\ref{fig:1}, they approach zero for large ${\bar r}$.}
	\end{figure}

In Fig.~\ref{fig:3}, the dependence of the physically relevant functions  $(\B_1+\B_2)\,\Omega^{-4}$ and $\B_1\Omega^{-4}$ on ${\bar r}$ is plotted. Interestingly, while the former approaches zero monotonously, the latter has its extreme at a specific distance outside the horizon ${\bar r}_h$. At this radius, the transverse Bachian tidal effect \eqref{InvGeoDevBHir0} is maximal.

\section{Thermodynamical properties: horizon area, temperature, entropy}

It is also important  to determine main physical properties
of the family of Schwarzschild--Bach black holes.
The \emph{horizon} in these spherically symmetric spacetimes is
generated by the (rescaled) null Killing vector ${\ell\equiv\sigma\partial_t=\sigma\partial_u}$ and
thus is located at ${\H=0}$, i.e., at ${r=r_h}$, see~\eqref{horizon}, \eqref{H_[0,1]}.
Its \emph{area} is, using  \eqref{Omega_[0,1]},
\be
{\cal A} = 4\pi\,{\bar r}_h^2 = 4\pi\,\Omega^2(r_h)= 4\pi\,r_h^{\,-2}\,.
\label{horizon_area}
\ee

Non-zero derivatives of $\ell$ are
${\ell_{u;r}=-\ell_{r;u}=\frac{1}{2}\sigma(\Omega^2\H)'}$. The
\emph{surface gravity}, given by
${\kappa^2\equiv-\frac{1}{2}\,\ell_{\mu;\nu}\,\ell^{\,\mu;\nu}}$
on the horizon \cite{Wald:1984}, is thus
\be
\kappa/\sigma
= -\tfrac{1}{2}\,\H'(r_h) = -\tfrac{1}{2}\,r_h =\tfrac{1}{2}\,
{\bar r}_h^{\,-1} \,.
\label{surface_gravity}
\ee
It is \emph{the same expression as in the Schwarzschild case}
(${\sigma=1}$, ${\kappa=\frac{1}{4m}}$), independent of the
Bach parameter $b$. The value of the scaling factor $\sigma$ is
fixed by the condition that ${h=-\Omega^2\H\to1}$
asymptotically as ${\bar{r}=\Omega(r)\to\infty}$.

The black-hole horizon \emph{temperature} is thus
\be
T/\sigma = \tfrac{1}{2\pi}\,\kappa/\sigma = -\tfrac{1}{4\pi}\,r_h
  = \tfrac{1}{4\pi}\,{\bar r}_h^{\,-1} \,.
\label{temperature}
\ee

However, in higher-derivative theories we have to apply the generalized
definition of \emph{entropy}
${S=(2\pi/\kappa)\oint \mathbf{Q}\,}$,  see \cite{Wald:1993}, where the Noether charge 2-form on the horizon is
\be
\mathbf{Q} =
-\frac{\Omega^2\, \H' }{16\pi}
\Big[1+\tfrac{4}{3}k^2\,\frac{\B_1+\B_2}{\Omega^4}\Big]\Big|_{r=r_h}\!
\sin\theta\,\dd\theta\wedge\dd\phi \,.
\label{Noether}
\ee
Evaluating the integral, using \eqref{horizon_area},
\eqref{surface_gravity}, \eqref{bonhorizon}, we get
\be
S = \tfrac{1}{4}{\cal A}\,\big(1-4k\,r_h^2\,b\big)
  = \tfrac{1}{4}{\cal A}\,\big(1-4k\,{\bar r}_h^{\,-2}\,b\big) \,.
\label{entropy}
\ee
This explicit formula for the Schwa--Bach
black hole entropy agrees with the results of
\cite{LuPerkinsPopeStelle:2015}, with the identification
${k=\alpha}$, ${b=\delta^*}$. In fact, it gives a \emph{geometric
interpretation} of the ``non-Schwarzschild parameter''
$\delta$  as the parameter $b$
determining the \emph{value of the Bach tensor on the horizon}, see \eqref{bonhorizon}.
 For the Schwarzschild black hole (${b=0}$) or
in Einstein's theory (${k=0}$), we recover the standard expression. For smaller Schwa--Bach black holes (smaller ${\bar r}_h$),
the deviations from ${S=\frac{1}{4}{\cal A}}$ are larger, analogously to \cite{NozawaMaeda:2008}.
To retain ${S>0}$, it is necessary to have ${4kb<{\bar r}_h^2}$,  restricting the theory parameters if~${{\bar r}_h \to 0}$.

Combining expressions \eqref{temperature},\ \eqref{entropy},\ \eqref{horizon_area},
\emph{exact relation between the temperature and the entropy} is obtained,

\be
T = \tfrac{1}{4}\sigma\, \big(\pi\, S + 4 \pi^2 k\,b \big)^{-1/2} \,,
\label{TonS}
\ee
generalizing  ${T = \tfrac{1}{4} (\pi S )^{-1/2} }$ of the Schwarzschild case.

For the parameters of Fig.~\ref{fig:1}, the values of $S$ and
$T$ agree with those given by Eq.~(11) in
\cite{LuPerkinsPopeStelle:2015}. From the behavior of the
metric functions for large $\bar r$ we were also able to
estimate the \emph{mass of this Schwa--Bach black hole} as ${2M
\approx 0.55}$, also in full agreement with
\cite{LuPerkinsPopeStelle:2015}. In fact, in
\cite{LuPerkinsPopeStelle:2015} the mass for this whole family
of black holes was studied numerically, and the first law of
thermodynamics was confirmed.

Our current research topics are Schwarzschild--Bach black holes
with a cosmological constant, and the study of specific
astrophysical consequences (e.g., pericenter precession or
gravitational lensing).

\section{Acknowledgements}

We acknowledge the support by the Czech Science Foundation Grant No.
GA\v{C}R 17-01625S  (JP, R\v{S}) and the Research Plan RVO: 67985840
(VP, AP). We are also grateful to H.~Maeda, M.~Ortaggio, R.~Steinbauer and C.~S\"amann  for helpful comments.
 We thank H.~L\"u for the \textsc{Mathematica} code for numerical integration of the field equations, plotted in Fig.~\ref{fig:2}, and for the specific value of the parameter ${\delta^*}$.

\end{document}